\documentclass[11pt,twoside]{article}

\usepackage{asp2006}
\usepackage{epsf}
\usepackage{lscape}
\usepackage{graphicx} 

\begin{document}
\title{Circular Polarimetry of Extragalactic Radio Sources}   
\author{Elena Cenacchi,\altaffilmark{1} Alexander Kraus and Thomas Beckert}   
\affil{Max-Planck-Institut f\"ur Radioastronomie, Auf dem H\"ugel 69, D-53121 Bonn, Germany}    
\altaffiltext{1}{E.C. is a member of the International Max-Planck Research School for Radio and Infrared Astronomy}
\author{Karl-Heinz Mack}   
\affil{Istituto di Radioastronomia, INAF, Via P. Gobetti 101, I-40129 Bologna, Italy}    

\begin{abstract}
We report multi-frequency circular polarization measurements for the four extragalactic radio sources 0056-00, 0716+71, 3C138 and 3C161 taken at the Effelsberg 100-m radiotelescope. The data reduction is based on a new calibration procedure that allows the contemporary measurement of the four Stokes parameters at different frequencies with single-dish radiotelescopes. We are in the process of framing the observed full Stokes spectra within a theoretical model that explains that the level of measured circular polarization as Faraday conversion. 
\end{abstract}

\vspace{-0.15cm}
\begin{figure}[!t]
\label{p:0056-00}
\plotone{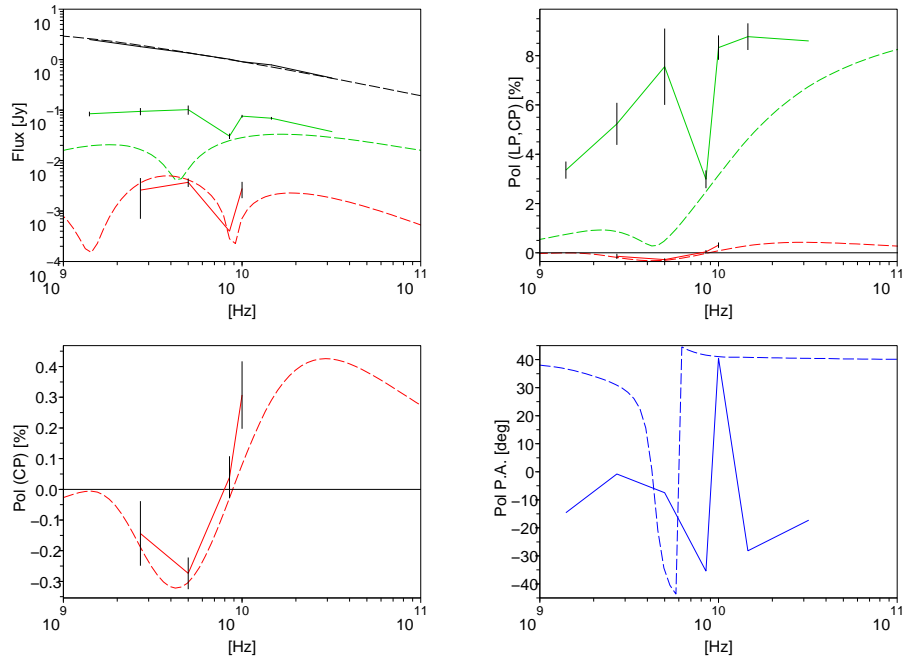}
\caption{0056-00, each plot contains a comparison of the observed radio continuum spectra (continuum line) and the simulated one (dashed line). Upper left, top to bottom: flux density I (the predicted values are almost coincident with the observed one), LP, CP (absolute value). Upper right, top to bottom: LP, CP. Lower left: CP. Lower right: PA.}
\end{figure}

\section{Introduction}   
The production of circular polarization (CP) in radio sources is typically ascribed either to intrinsic synchrotron emission or to Faraday conversion: a birefringence effect that converts linear polarization (LP) into CP \citep{WardleHoman2003}. In a relativistic plasma the normal modes are elliptical and both Faraday rotation and conversion are present. Theoretically, the conversion due to propagation should dominate any contribution from intrinsic emission, however so far there are only few observational confirmations. 

The multi-frequency full Stokes polarimetry is a powerful tool to study the radiation emission and transfer processes in the observed sources and 
determine the dominant mechanism for circular polarization production. CP, LP and spectral information can be used to constrain the low energy end of the relativistic particle distribution \citep{Beckert}, derive magnetic field order, strength and geometry \citep{Gabuzda} and make assumptions about the composition of the relativistic plasma within jets \citep{Wardle}. 

Measuring CP with a single-dish radiotelescope is a challenging task, due to the generally small values involved and the high accuracy required. We have developed a new procedure to calibrate full-Stokes polarimetric data obtained with the Effelsberg 100-m radiotelescope \citep{Cenacchi} which allows a complete description of the instrumental M\"uller matrix and its correction. We have tested it nearly monthly at 5 GHz on a sample of 43 sources during 2007. In 2008 we have applied our procedure to 2.7, 8.5 and 10-GHz data of a sub-sample of four sources that systematically exhibited a level of CP above 3$\sigma$ and fractional LP above 1\% at all frequencies.

\section{Observations}
The instrumental M\"{u}ller matrix elements contain information about the amount of contamination among the four Stokes parameters, due to spurious, unwanted, conversions arisen inside the receiver. This clearly appears when dealing with completely unpolarized (thermal) sources (e.g. the planetary nebulae NGC7027 and NGC6572 show levels of CP and LP of nearly 0.5\% before the M\"uller matrix correction). 

The sources were observed with 5-beam wide cross-scans centered on the source. The $Q$ and $U$ values come directly from the polarimeter. The $V$ value has been estimated as difference of the left-hand circular and right-hand circular channels, after that these had been separately corrected for pointing (to compensate for the beam squint) and gain curve. The cross-scan technique intrinsically removes possible differences between the internal channel noises. The correction for the 4$\times$4 M\"uller matrix has been applied by solving the equations for the leackage terms \citep{Cenacchi} using NGC7027 as completely unpolarized calibrator and 3C286 as strongly linearly polarized one (no assumption about Stokes $V$ from the linearly polarized calibrator is required). 

\vspace{-0.15cm}
\begin{figure}[!t]
\label{p:0716+71}
\plotone{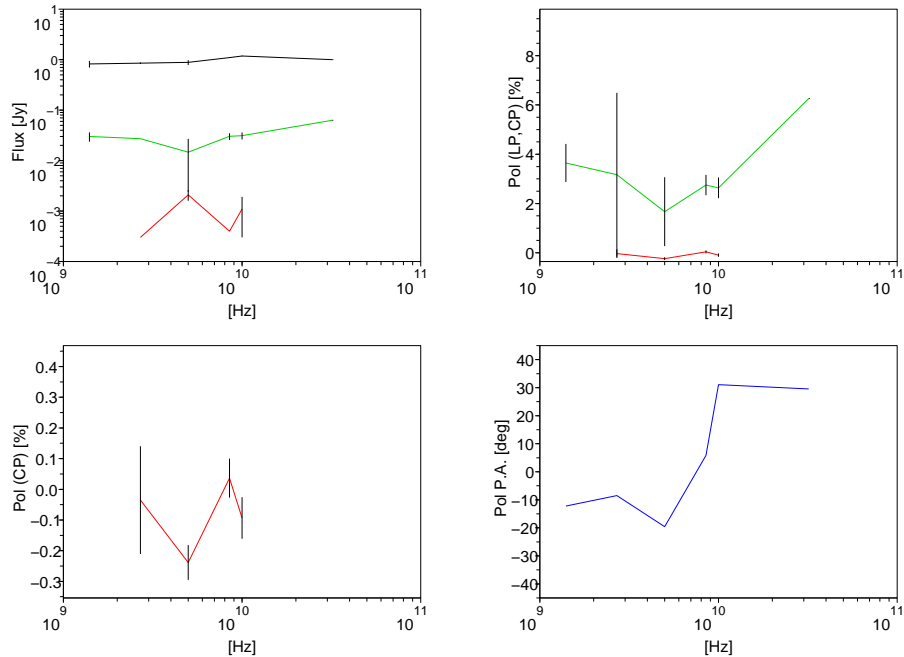}
\caption{0716+71, observed spectra. Upper left, top to bottom: flux density I, LP, CP (absolute value). Upper right, top to bottom: LP, CP. Lower left: CP. Lower right: PA.}
\end{figure}

\vspace{-0.15cm}
\begin{figure}[!t]
\label{p:3C138}
\plotone{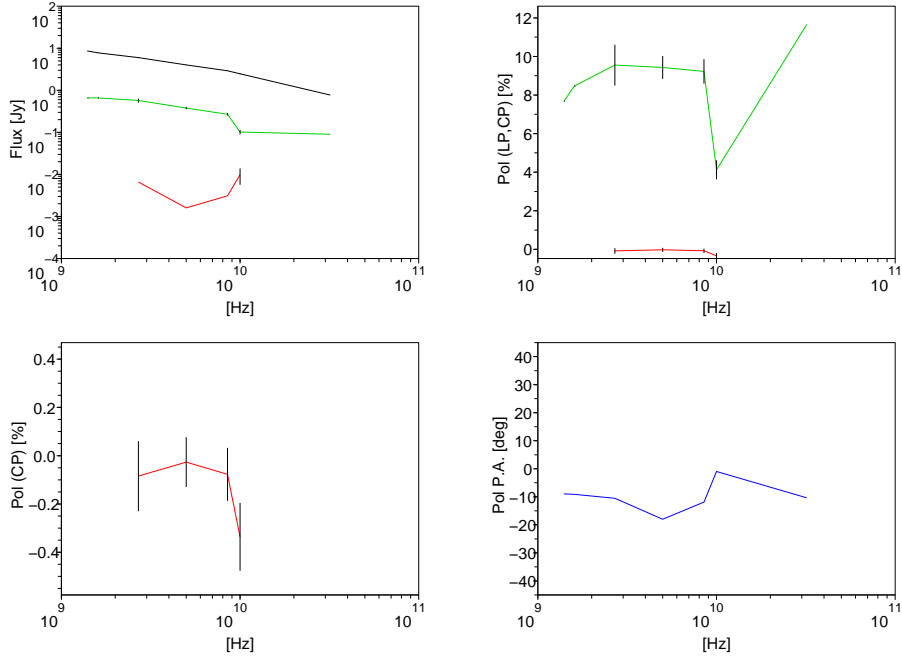}
\caption{3C138, observed spectra. Upper left, top to bottom: flux density I, LP, CP (absolute value). Upper right, top to bottom: LP, CP. Lower left: CP. Lower right: PA.}
\end{figure}

\section{Results and Discussion}
\textbf{0056-00}: This source exhibits a change in sign of CP between 5 and 8.5 GHz, consistent with the minimum in LP and with the change in polarization angle. We have compared our results with those obtained with a model developed by Beckert \& Falcke (2002) that provides the radiative transfer coefficients for polarized synchrotron radiation applied to the standard model for relativistic radio jets. The model assumes an extended unpolarized synchrotron source, which dominates the flux density below nearly 5 GHz which is modeled with energy equipartition between B-field and particles, $B$=4 mG, $n_e=10^{-3}$ cm$^{-3}$, and  a typical power-law for electrons above $\gamma$=100, with $p$=2.45 (power-law index). The size of the emitting region is $L=8 \cdot 10^{21}$ cm. The polarized emission would be produced by a compact jet component of $L=1.5\cdot 10^{19}$ cm, $B$=90 mG, $n_e$=0.5 cm$^{-3}$ with a well-ordered magnetic field (a tightly wound spiral) seen at an angle of 85$^\circ$. This component becomes self-absorbed below 6 GHz and the emission is relativistically boosted with $\Gamma$=6. This combination reproduces the observed level of circular polarization, the sign flip at nearly 8 GHz and the observed flux density. Within the current model the minimum in LP is at lower frequencies and the simulated level of LP is too low which indicates that there might be an additional component (even more compact) that dominates LP and produces the turn in polarization angle at higher frequencies. The observed levels of CP are: $-0.14\pm 0.11\%$ at 2.8 GHz, $-0.28\pm 0.05\%$ at 5 GHz, $0.04\pm0.07\%$ at 8.5 GHz, $0.31\pm 0.11\%$ at 10.GHz.

\textbf{0716+71}: This BLLAC source exhibits a level of CP of: $0.03\pm 0.17\%$ at 2.7 GHz, $-0.24\pm 0.06\%$ at 5 GHz, $0.04\pm0.06\%$ at 8.5 GHz, $-0.09\pm 0.07\%$ at 10.5 GHz and the minimum in CP is aligned with the minimum in LP and with the turn in polarization angle.

\textbf{3C138}: Observed CP values for the quasar 3C138 are: $-0.11\pm 0.19\%$ at 2.7 GHz, $-0.04\pm 0.15\%$ at 5 GHz, $-0.11\pm0.15\%$ at 8.5 GHz, $-0.40\pm 0.17\%$ at 10.5 GHz  Also in this case the minimum in LP is aligned with a drop in CP and with a turn in polarization angle. Possible future CP measurements at higher frequencies could be helpful to identify the exact location of the CP minimum.

\textbf{3C161}: The quasar 3C161 presents a full Stokes spectrum that differs from the others and appears more complicated as explained by our model. Observed CP values are $0.08\pm 0.16\%$ at 2.7 GHz, $-0.20\pm 0.08\%$ at 5 GHz, $0.09\pm0.06\%$ at 8.5 GHz, $0.28\pm 0.12\%$ at 10.5 GHz. 

\vspace{-0.15cm}
\begin{figure}[!t]
\label{p:3C161}
\plotone{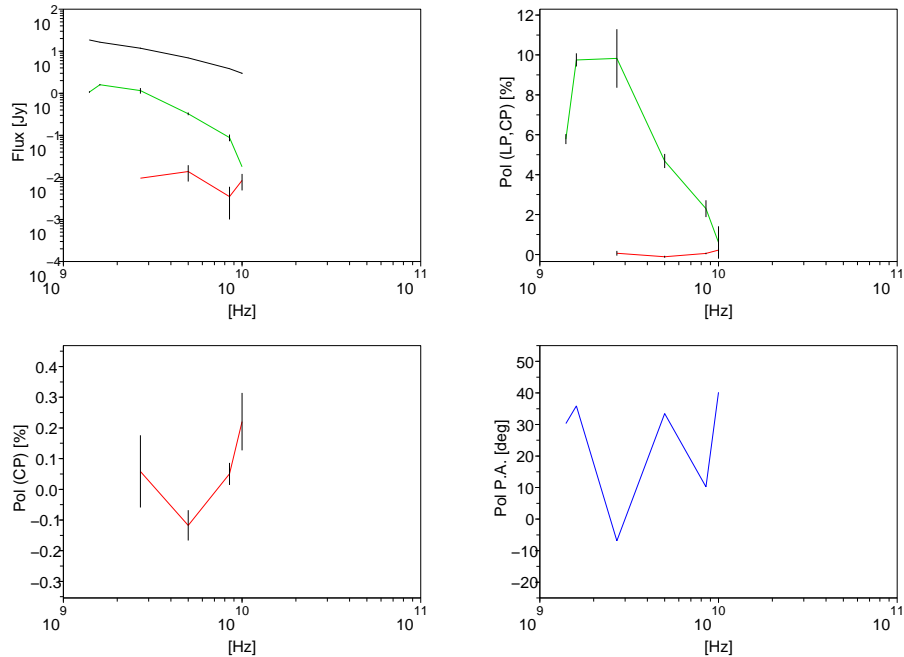}
\caption{3C161, observed spectra. Upper left, top to bottom: flux density I, LP, CP (absolute value). Upper right, top to bottom: LP, CP. Lower left: CP. Lower right: PA.}
\end{figure}

\section{Conclusions}
We have developed a new polarimetric calibration procedure that allows multi-frequency full Stokes observations and can be generalized to any telescope equip\-ped with circular outputs and radio polarimeters. After extensive testing carried out on 43 sources at 5 GHz, we have applied the calibration at 2.7, 8.5 and 10 GHz obtaining the full Stokes spectra of those four sources characterized by a level of CP above 3$\sigma$ at 6 cm. We are comparing the observed results with those simulated by a model that assumes the Faraday conversion as origin for the observed CP. The values observed in 0056-00 are consistent with those predicted by the model. A similar behavior seems to be indicated by preliminary studies on 3C138 and 0716+71 while the behavior of 3C161 is more difficult to explain and needs more thorough analysis. Further full Stokes measurements are planned at 14.5 and 32 GHz.

\acknowledgements 
This research was supported by the EU Framework 6 Marie Curie Early Stage Training programme under contract number MEST-CT-2005-19669 "ESTRELA".

\end{document}